\documentclass{aa}
\usepackage{graphicx}

\begin{document}

\title{The Calar Alto Deep Imaging Survey:\\
       $K$-band Galaxy Number Counts
       \thanks{Based on observations collected at the German-Spanish 
               Astronomical Centre, Calar Alto, operated by the 
               Max-Planck-Institut f\"ur Astronomie, Heidelberg, jointly with 
               the Spanish National Commission for Astronomy}}
\titlerunning{The CADIS near-infrared and optical number counts}
\authorrunning{J.-S. Huang et al.}

\author{J.-S. Huang\inst{2,}\inst{1}, 
        D. Thompson\inst{3,}\inst{1},
        M.\ W. K\"ummel\inst{1}, 
        K. Meisenheimer\inst{1},
        C. Wolf\inst{1}, 
        S.\ V.\ W. Beckwith\inst{4,}\inst{1},
        R. Fockenbrock\inst{1},
        J.\ W. Fried\inst{1},
        H. Hippelein\inst{1},
        B. von Kuhlmann\inst{1},
        S. Phleps\inst{1},
        H.-J. R\"oser\inst{1}
    and E. Thommes\inst{1}}
\institute{Max-Planck Institut f{\"u}r Astronomie,
           K{\"o}nigstuhl 17, D-69117 Heidelberg, Germany
      \and (present address:) 
           Harvard-Smithsonian Center for Astrophysics,
           60 Garden Street, 
           Cambridge, MA  02138, USA
      \and (present address:) 
           California Institute of Technology, MS 320-47,
           Pasadena, CA  91125, USA
      \and (present address:) 
           STScI, 3700 San Martin Dr., 
           Baltimore, MD  21218, USA}

\offprints{J.-S. Huang}
\mail{jhuang@cfa.harvard.edu}
\date{Received 13 April 2000/ Accepted 9 January 2001}

\abstract{
We present $K$-band number counts for the faint galaxies in 
the Calar Alto Deep Imaging Survey (CADIS).  We covered 4 CADIS fields, 
a total area of $0.2\,\mathrm{deg}^2$, in the broad band filters $B$, 
$R$ and $K$. We detect about 4000 galaxies in the $K$-band images, with 
a completeness limit of $K=19.75\,\mathrm{mag}$, and derive 
the $K$-band galaxy number counts in the range of $14.25 < K < 
19.75\,\mathrm{mag}$.  This is the largest medium deep $K$-band survey 
to date in this magnitude range.  The $B$- and $R$-band number counts 
are also derived, down to completeness limits of $B=24.75\,\mathrm{mag}$ 
and $R=23.25\,\mathrm{mag}$.  The $K$-selected galaxies in this magnitude 
range are of particular interest, since some medium deep near-infrared 
surveys have identified breaks of both the slope of the $K$-band number 
counts and the mean $B-K$ color at $K=17\sim18\,\mathrm{mag}$.  There is, 
however, a significant disagreement in the $K$-band  number counts among 
the existing surveys.  Our large near-infrared selected galaxy sample
allows us to establish the presence of a clear break in the slope at
$K=17.0\,\mathrm{mag}$ from $\mathrm{d}\log N/\mathrm{d}m = 0.64$ at 
brighter magnitudes to $\mathrm{d}\log N/\mathrm{d}m = 0.36$ at the fainter 
end.  We construct no-evolution and passive evolution models, and find
that the passive evolution model can simultaneously fit the $B$-, $R$- and 
$K$-band number counts well.  The $B-K$ colors show a clear trend to bluer 
colors for $K > 18\,\mathrm{mag}$.  We also find that most of the 
$K=18-20\,\mathrm{mag}$ galaxies have a $B-K$ color bluer than the 
prediction of a no-evolution model for an $L_*$ Sbc galaxy, implying either 
significant evolution, even for massive galaxies, or the existence of an 
extra population of small galaxies. 
}

\maketitle


\section{Introduction} \label{sec:intro}

Many galaxy surveys have shown that galaxies were undergoing a significant
evolution at intermediate redshifts ($0.5<z<1.0$).  Cowie et al. 
(\cite{cow96}) and Ellis (\cite{ell97}, et al.~\cite{ell96}) suggested that 
an extra galaxy population in this redshift range is responsible for the 
excess of the galaxy number counts at faint magnitudes.  With additional 
morphological classification for the faint galaxies, this extra population 
was identified as irregular and peculiar galaxies with strong star formation 
at intermediate redshifts (Glazebrook et al.~\cite{gla95}, Abraham et 
al.~\cite{abr96}, van den Bergh et al.~\cite{vdb96}, Huang et 
al.~\cite{hua98}).  The massive elliptical and giant spiral galaxies,
however, show no significant evidence of evolution since $z=1.0$
(Lilly et al.~\cite{lil95}).

As an indication that the faint blue galaxy excess has a significant
contribution from galaxies at high redshift
Cowie et al. (\cite{cow97}) find that at least some of their faint 
$B$-selected galaxies are located in a high-redshift ($1<z<2$) tail 
on the redshift distribution. 
Dickinson (\cite{dic00}) conducted a 
near-infrared study on HDF-N galaxies at using the NICMOS camera on HST 
and found far fewer high luminosity galaxies in the $1.4 < z < 2$ range.   
However, the NICMOS data cover only a small field, and may not be a 
representative sample.  Steidel et al. (\cite{ste96},\cite{ste99}) have 
identified a large number of galaxies at higher redshifts ($z\sim3-4$) 
using the Lyman-break technique.  However, these galaxies do not contribute 
significantly to the faint blue galaxy number counts at $B < 
26\,\mathrm{mag}$.  For $z > 3$, the $B$ band covers wavelengths below 
the redshifted Lyman-alpha ($\lambda_{\mathrm{rest}}=1216\,$\AA) 
line, where the rest-frame UV light is strongly affected by absorption in 
the Ly-$\alpha$ forest as well as dust. 

Comparing the galaxy luminosity function at different redshifts is the 
most direct way of understanding the galaxy evolution.  However existing 
samples are not large enough and the galaxy luminosity function at 
intermediate redshifts is not well determined (Lilly et al.~\cite{lil95}, 
Cowie et al.~\cite{cow96}, Ellis et al.~\cite{ell96}), especially at lower 
luminosities.  A wide field medium deep survey is required to obtain larger 
numbers of low luminosity galaxies, to better constrain the luminosity 
function at intermediate redshift.  While large-format CCDs have been 
in use for some time now, sufficiently large near-infrared arrays 
have only recently become available.

There are three significant advantages to studying galaxy evolution in 
the near-infrared $K$ band instead of in the optical bands (Cowie et 
al.~\cite{cow94}, Huang et al.~\cite{hua97}).  First, the cosmological 
k-correction in the $K$-band is much smaller and better determined 
than the k-corrections in the optical bands.  Second, the shorter 
wavelength optical bands sample restframe blue or UV light, and are 
thus very sensitive to the star-formation history in galaxies.  
Third, the $K$-band light is a close tracer of the stellar mass in 
galaxies, and less sensitive to either the star formation history or 
the galaxy spectral type.  A $K$-selected sample of galaxies is thus more 
representative of the true distribution of galaxies.  

A number of medium deep and deep $K$-band galaxy surveys have been conducted 
by various groups (Gardner et al.~\cite{gar93}, Songaila et al.~\cite{son94},
Djorgovski et al.\ \cite{djo95}, Saracco et al.~\cite{sar97,sar99},
Minezaki et al.~\cite{min98}, Bershady et al.~\cite{bers}, Moustakas et 
al.~\cite{mou98} and K\"ummel \& Wagner \cite{kue}).  There is some disagreement between the $K$-band number 
counts of these surveys at faint magnitudes.
Still, these surveys do show evidence for two critical phenomena which are 
important in understanding galaxy evolution: the mean $B-K$ color becomes 
bluer at faint magnitudes, and the slope of the $K$-band number counts shows 
a break in the $17 < K < 19$ magnitude range.  The Hawaii $K$-band galaxy
survey (Songaila et al.~\cite{son94}) 
shows that galaxies in the range of $16 < K < 20$ magnitude are located at 
$0.3<z<1.3$, which spans an important epoch for galaxy evolution. 

%
\begin{table}
   \caption{Field coordinates \label{tbl:1}}
   \begin{tabular}{lccc}
      \hline
      \noalign{\smallskip}
      Field& $\alpha$(2000)&$\delta$(2000) & area [$\mathrm{arcmin}^2$]\\
      \noalign{\smallskip}
      \hline
      \noalign{\smallskip}
      01H & 01:47:33.3 & 02:19:55 & 175\\
      09H & 09:13:47.5 & 46:14:20 & 127\\
      16H & 16:24:32.3 & 55:44:32 & 186\\
      23H & 23:15:46.9 & 11:17:00 & 188\\
      \noalign{\smallskip}
      \hline
   \end{tabular}
\end{table}

To better determine the $K$-band number counts in this magnitude range, a
$K$-band survey with much larger coverage is essential.
Here, we use the Calar Alto Deep Imaging Survey (CADIS, Meisenheimer et 
al.~\cite{mei99}) to derive improved $K$-band number counts in the range 
$14.25 < K < 19.75\,\mathrm{mag}$, as well as the corresponding optical 
number counts in the $B$ and $R$ bands.  CADIS was originally tailored 
to search for Ly-$\alpha$ emitters at high redshift with Fabry-Perot 
Interferometer (FPI) imaging (Thommes et al.~\cite{tho98}).  However, 
the CADIS fields are also observed in three broad band filters ($B$, $R$, 
$J$ or $K$) and up to $13$ medium band filters with $\lambda /\Delta 
\lambda = 25-50$.  This multi-color database is also being used to study 
galaxy populations and their evolution at intermediate redshifts through 
their photometric redshifts (Wolf \cite{wo98}), as well as the detection 
of unusual objects like active galactic nuclei (Wolf et al.~\cite{wo99}) 
and extremely red galaxies (Thompson et al.~\cite{tho99}).  Fried et 
al.~(\cite{fried}) present initial results from CADIS on the luminosity 
function of optically selected galaxies and their evolution.  

In this paper, we present the optical and near-infrared galaxy number 
counts, as well as the colors of the $K$-selected galaxies.  The four 
CADIS fields used for this study cover a total of 0.2 deg$^2$ down to 
a completeness limit of $K=19.75\,\mathrm{mag}$.  Our coverage is four 
times the area of the ESO $K$-band survey (Saracco et al.~\cite{sar97}), 
which is the largest medium deep $K$-band survey to date.  The paper is 
organized as follows: The observation, data reduction and photometry are 
described in Section 2; the $B$-, $R$-, and $K$-band number counts are 
presented in Section 3, along with the colors of the $K$-selected galaxies.  
In Section 4 we present a galaxy model which reproduces the observed number 
counts, and Section 5 summarizes our results.

\section{Observation and photometry} \label{sec:obs}

\subsection{Observation and data reduction}

%
\begin{table*}
   \centering
   \caption{Observational parameters \label{tbl:2}}
   \begin{tabular}{lcrccrccrcrc}
   \hline
   \noalign{\smallskip}
                                         & 
      \multicolumn{3}{c}{B}              & 
      \multicolumn{3}{c}{R}              &
      \multicolumn{2}{c}{I\footnotemark} & 
      \multicolumn{3}{c}{K}              \\
      Field & Tel\footnotemark & Int\footnotemark & M$_{90}$\footnotemark & 
              Tel              & Int              & M$_{90}$              & 
              Tel              & Int              & 
              Tel              & Int              & M$_{90}$              \\
   \noalign{\smallskip}
   \hline
   \noalign{\smallskip}
      01H & 2.2& 5050& 23.7& 2.2& 2800& 22.8& 2.2&  4000& 3.5&  9648& 19.4 \\
      09H & 3.5& 4000& 24.9& 2.2& 2800& 23.4& 3.5&  7000& 3.5& 11280& 18.8 \\
      16H & 2.2& 6200& 24.6& 2.2& 2500& 23.7& 2.2& 13620& 3.5&  9067& 19.3 \\
      23H & 2.2& 6230& 23.4& 2.2& 2500& 23.0& 2.2&  5000& 3.5&  6475& 19.6 \\
   \noalign{\smallskip}
   \hline
   \end{tabular}
\end{table*}
\addtocounter{footnote}{-4}
\stepcounter{footnote}
\footnotetext{The CADIS $I$ filter used here is a narrow band filter at
              $\lambda = 815\,\mathrm{nm}$ ($\delta \lambda =25\,\mathrm{nm}$),
              and used only for the star-galaxy classification.  The object 
              detection program was not run on the I-band images.}
\stepcounter{footnote}
\footnotetext{The Calar Alto 2.2\,m and 3.5\,m telescopes.}
\stepcounter{footnote}
\footnotetext{The integration time in seconds.}
\stepcounter{footnote}
\footnotetext{M$_{90}$ is the magnitude where 90\% of the galaxies are 
              detected in our Monte Carlo test.}
  
A complete description of the observation and data reduction for CADIS will 
be given elsewhere (Meisenheimer et al. in prep.), here we give only a 
brief description.  The CADIS fields are selected at high galactic latitudes 
on the northern and equatorial sky.  The two primary CADIS field selection 
criteria are: (1) no bright star ($R<16\,\mathrm{mag}$) within a field of 
12'$\times$12', and (2) a local minimum on the IRAS $100\,\mu m$ maps, with 
an absolute surface brightness of $<2\,\mathrm{MJy/sr}$.  The field positions 
and areas covered in this project are given in Table~\ref{tbl:1}.  

The optical observations were obtained with focal reducing CCD cameras 
on the 2.2m and the 3.5m telescope on Calar Alto Spain.
The {\it Calar Alto Faint Object Spectrograph}
(CAFOS) covers a 200$\,\mathrm{arcmin}^2$ field at 0.53 arcsec per pixel 
on the 2.2\,m telescope, while the {\it Multi-Object Spectrograph 
for Calar Alto} (MOSCA) covers a $150\,\mathrm{arcmin}^2$ at 0.33 arcsec 
per pixel on the 3.5\,m telescope.  Only images with seeing better than 
$1\farcs 8$ are included in the CADIS database.  

The $K$-band data were obtained with the Omega-Prime near-infrared camera 
(Bizenberger et al.~\cite{biz98}) on the 3.5\,m telescope at Calar Alto.  Omega-Prime 
has a 1024$\times$1024 pixel HgCdTe array which, at 0.396 arcsec per pixel, 
covers a $6.8'\times 6.8'$ field of view.  This camera has no cold pupil 
stop, which would effectively block thermal emission from the warm surface 
around the primary mirror.  We used a $K'$-filter (Wainscoat \& Cowie 
\cite{wai92}) instead of the standard $K$-band to reduce the thermal 
background.  In order to construct a larger field of view, matching that
of the optical cameras, we observed the CADIS-fields in a $2 \times 2$ 
mosaic.  After trimming the high noise edges of the stacked mosaics, the 
final coverage for the $K$-band images varies from $127\,\mathrm{arcmin}^2$ 
to $188\,\mathrm{arcmin}^2$ (see Tables~\ref{tbl:1} and \ref{tbl:2}).
The $K$ band image quality criterion is a seeing less than $1\farcs 5$.

The optical images were initially  processed using the Munich Image Data
Analysis System (MIDAS) software package, while the near-infrared images
were processed using the Interactive Data Language (IDL) or IRAF\footnotemark.
\footnotetext{IRAF is distributed by the National Optical Astronomy
Observatories, which are operated by the Association of Universities for
Research in Astronomy, Inc., under cooperative agreement with the National
Science Foundation}
Thompson et al.~(\cite{tho99}) gives a discussion of the
steps of the standard reduction to get the final, stacked images in the
optical and the near-infrared.  A detailed discussion on the detectability 
and photometry is given in the next section.

Photometric calibrations for all of the CADIS data are determined on 
the Vega scale.  Each CADIS field has one or two relatively bright stars 
for which we have established accurate spectro-photometric calibrations 
between $350\,\mathrm{nm}$ and $950\,\mathrm{nm}$.  The magnitudes of these 
tertiary standards are then obtained by integrating their spectral energy 
distributions within the CADIS filter profiles.  Wolf~(\cite{wo98}) gives 
a detailed description of the CADIS magnitude system and its calibration.  
For the $K$-band data, we use the UKIRT faint standard stars (Casali \& 
Hawarden \cite{cas92}) for flux calibration.  The total exposure times and 
detection limits for the optical and infrared data are given in 
Table~\ref{tbl:2}.  

We use the CADIS broad band $B$, $R$, and $K$ data for the number counts, 
and add the medium band $I$ ($\delta \lambda = 25\,\mathrm{nm}$, centered 
at $815\,\mathrm{nm}$) data in order to facilitate star-galaxy separation. 
The CADIS B and R filters are very close to the Johnson-Cousins system
($B_J$ and $R_{kc}$ in the equations below).  The transformations between 
the two systems were derived using synthetic photometry of the galaxy 
templates of Kinney et al.~(\cite{kin}) in the redshift range $0<z<1.5$.  The 
transformation equations are:
\begin{equation}
   B_{J}-B_{\mathrm{CADIS}}=0.082+0.121(B-R)_{\mathrm{CADIS}}
\end{equation}
\begin{equation}
   R_{kc}-R_{\mathrm{CADIS}}=-0.027-0.034(B-R)_{\mathrm{CADIS}}
\end{equation}

\subsection{Object detection and photometry} \label{sec:det}
 
Object catalogs were created using the SExtractor image analysis package 
(Bertin \& Arnouts \cite{ber96}).  Before running the detection program, 
each image was convolved with a Gaussian PSF with a width equal to the 
full width at half maximum (FWHM) of the seeing.  For detection, we 
required at least 5 connected pixels, each with a signal larger than three 
times the background noise.  Faint galaxy detectability is 
determined using Monte Carlo simulations.  Several bright galaxies with 
different morphological types were rescaled then added back into the original 
image before running the detection program again.  The result is the 
percentage of galaxies recovered as a function of apparent magnitude.  The 
magnitudes at which 90\% of the test galaxies are recovered are listed in 
Table~\ref{tbl:2}.

We used the photometry package MPIAPHOT (Meisenheimer \& 
R\"oser~\cite{mei86}), which was developed to measure accurate spectral 
energy distributions (SED) from images with variations in seeing.  In 
MPIAPHOT, the pixel values within a circular aperture are evaluated using 
a Gaussian weighting function.  The e-folding width $s_w$ of the Gaussian 
weighting function is set individually for each frame $i$ to
\begin{equation}
s_{w,i,x} = \sqrt{s_{\mathrm{eff}}^2 -s^2_{i,x}}.
\end{equation}
Here $s_{i}$ denotes the e-folding width of the seeing where $x=a,b$ 
corresponds to the major/minor axis of the empirical seeing profile and 
$s _{\mathrm{eff}}$ denotes the e-folding width of the {\it effective PSF}.
The effective seeing is chosen as a constant for the dataset of every
CADIS field, and compensates for variations in the seeing between images 
obtained at different times and through different filters.  Since most of 
our images have a seeing of about $1\farcs 5$ FWHM ($s_i = 0\farcs 90$) and 
the signal to noise ratio of the photometry has its maximum at $s_{\mathrm{eff}} \sim 
\sqrt{2} s_{i}$, the effective seeing is set to $s_{\mathrm{eff}}=1\farcs38$ 
for all data presented here.  

The weighted pixel values are summed within an aperture of $4''$ diameter, 
and then corrected to a total magnitude.  By default MPIAPHOT applies 
corrections calculated by integrating the flux of point-like objects to 
infinite radius.  The corrected magnitude thus equals the total magnitude 
for point sources, but underestimates the total magnitude of extended 
galaxies.  To correct for this effect, we apply another correction for 
galaxies.  The corrections are determined from simulations containing 
galaxies which match the range of apparent sizes seen in the survey data.
We find that the  corrections for resolved galaxies are almost independent 
of the bulge to disk ratios and inclinations.  The correction can be well 
described as a function of the measured size of the galaxies (i.e. the 
second moment along its major and minor axis) out to about $3''$ FWHM.  
The derived correction formula has been applied to the science data,
in order to convert CADIS magnitudes to total magnitudes.

\subsection{Star-Galaxy Separation}

%
\begin{figure}
   \centering
   \resizebox{6cm}{!}{\includegraphics*[angle=-0]{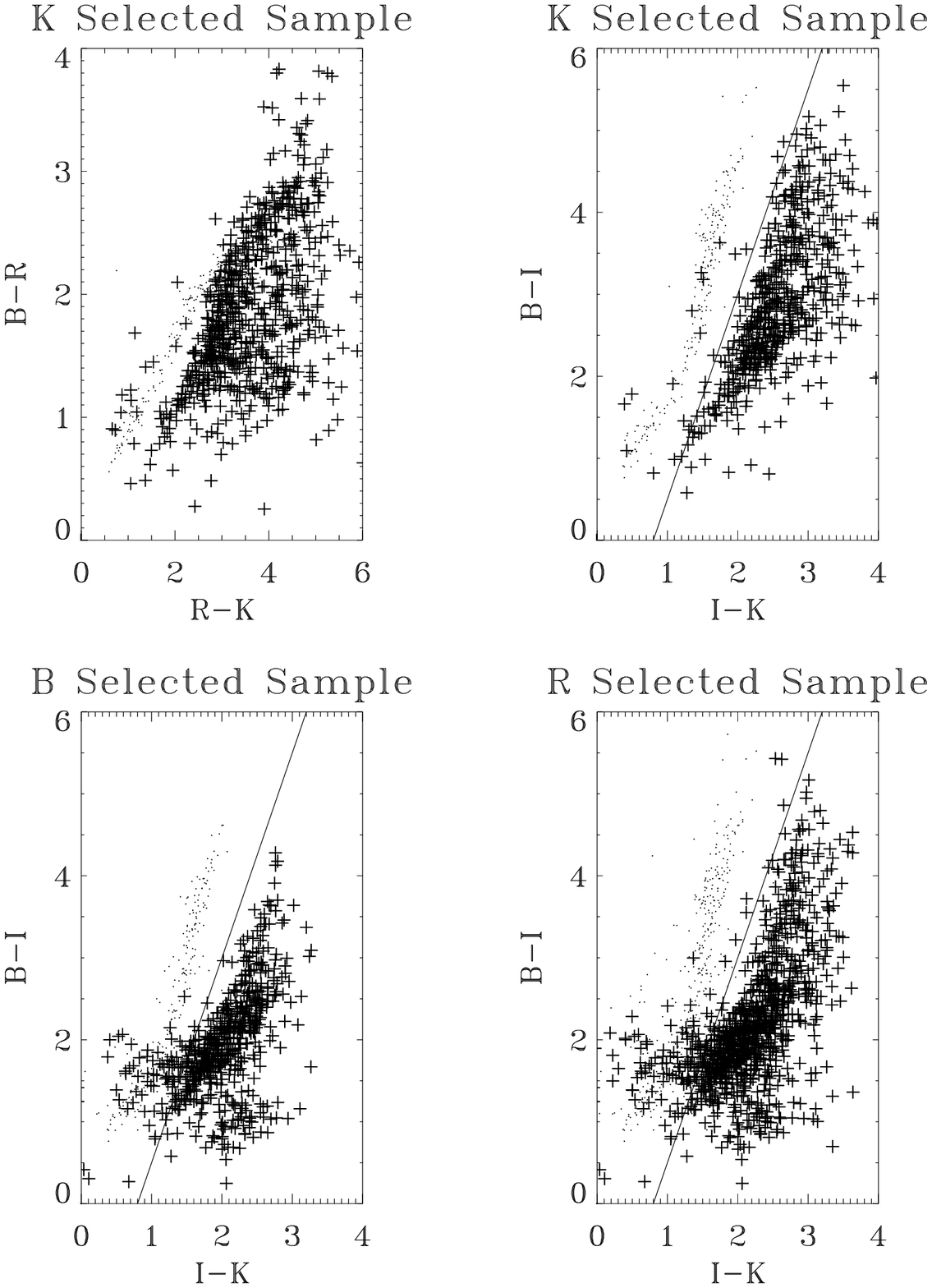}}
   \caption{The color-color diagram for the $K$-selected sample.  In the 
            diagram, the dots represent the point-like objects and the crosses 
            represent the extended objects.  The solid line represents the 
            empirical color criterion for star-galaxy separation (Cowie et 
            al.~\cite{cow94}, Huang et al.~\cite{hua97}).}
   \label{fig:2}
\end{figure}

Conventional star-galaxy separation is based on
comparing the morphology of objects with the morphology of point-like objects.
Since the $K$-images have the largest field size, there are regions where
only information in $K$ is available, and there we used this conventional
method for star-galaxy separation.
However, this method fails to identify compact galaxies with unresolved
morphology. While those objects are rare at bright magnitudes, this
population amounts to $\sim20\%$ of all galaxies
at $R=23\,\mathrm{mag}$. To identify this population
we use colors to separate stars and
galaxies. In three fields presented here (01h, 09h and 16h field)
the full CADIS dataset in broad bands and up to 13 medium band filters
is available. In those fields the broad wavelength coverage from
$0.4\,\mu\mathrm{m}$ to $2.2\,\mu\mathrm{m}$ allows a
classification into stars, galaxies and QSOs based on the object's SED.
This is done by comparing the measured color indices of every object with 
the expected color indices calculated from template spectra of the different
subclasses stars, galaxies and QSOs (Wolf \cite{wo98}).
For galaxies and QSOs the theoretical color indices are
calculated with a broad range of redshifts ($0<z<2$ for galaxies and $0<z<6$
for QSOs), the result of the classification
includes a redshift estimate with an accuracy of typically
$\delta z = 0.03$ for galaxies (Fried et al.\ \cite{fried}).
For galaxies, we use the set of templates published by
Kinney et al.~(\cite{kin}) whose SEDs range from an
elliptical to an extreme starburst galaxy. This further allows a classification
of the spectral type of galaxies found in the survey.
For the 23h-field, for which we do not yet have the full
color information, color-based classification was done using the color-color
diagram $I-K$ vs.\ $B-I$ (Cowie et al.~\cite{cow94}, Huang
et al.~\cite{hua97}). To demonstrate this technique we plot in 
Fig.\ \ref{fig:2} the color-color diagram $I-K$ vs.\ $B-I$
of pointlike (dots) and extended (crosses) objects for the $K$-selected
sample. In the $K$-selected sample stars and galaxies are well separated by
the  empirical relation
\begin{equation}
B-I=2.5*(I-K)-2
\end{equation}
with only a few galaxies crossing the line.

To summarize the criterion for the final star-galaxy separation:
an object is classified as a galaxy if {\it either} it is morphologically
identified as a galaxy, {\it or}
it is identified as a galaxy on the basis of its colors.

\section{Optical and near-infrared galaxy number counts}\label{sec:cou}
\begin{figure}
   \resizebox{\hsize}{!}{\includegraphics*[angle=-90]{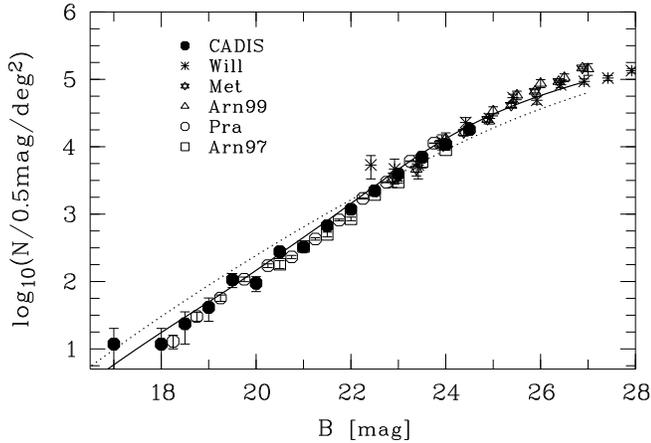}}
   \caption{The $B$-band galaxy number counts. The filled circles are the 
   number counts obtained in CADIS. The other symbols are number counts from
   Metcalfe et al.~(\cite{met95}) (Met), 
   Williams et al.~(\cite{wil}) (Will),
   Arnouts et al.~(\cite{arn97}) (Arn97),
   Prandoni et al.~(\cite{pra}) (Pra) and
   Arnouts et al.~(\cite{arn99}) (Arn99).
   The upper pair of lines show the passive evolution models, while the lower 
   pair of lines show the no evolution models.  In both cases, the solid lines 
   are for $q_0=0.05$ while the dashed lines are for $q_0=0.5$ models. }
   \label{fig:3}
\end{figure}

\begin{figure}
   \centering
   \resizebox{\hsize}{!}{\includegraphics*[angle=-90]{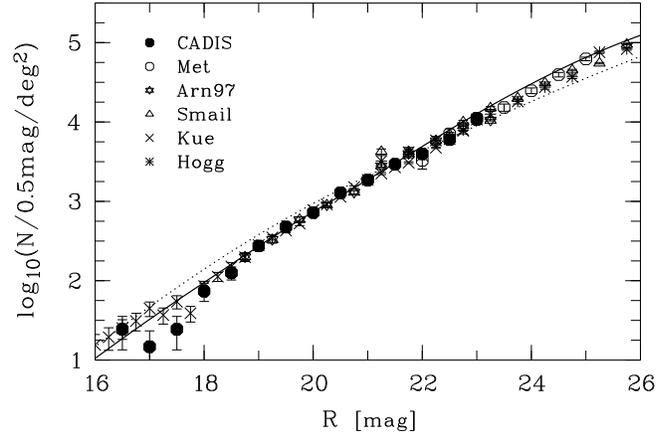}}
   \caption{The $R$-band galaxy number counts. The filled circles are the 
   number counts obtained in CADIS. The other symbols are number 
   counts from
   Metcalfe et al.~(\cite{met95}) (Met),
   Smail et al.~(\cite{smail}) (Smail),
   Arnouts et al.~(\cite{arn97}) (Arn97),
   Hogg et al.~(\cite{hogg}) (Hogg) and
   K\"ummel \& Wagner (\cite{kue2}) (Kue).
   The models for the $R$-band number counts are plotted as in 
   Fig.~\ref{fig:3}.}
   \label{fig:4}
\end{figure}

The galaxy number counts in the three filters are listed in 
Tables~\ref{tbl:3}-\ref{tbl:5}.  The number counts in $B$ and $R$ are 
plotted in Figs.\ \ref{fig:3} and \ref{fig:4}, respectively.  We also 
compare our number counts to those of other large scale surveys with areas 
on a one square degree scale, like Arnouts et al.~(\cite{arn97}), EIS
(ESO Imaging Survey, Prandoni et al.~\cite{pra}) and K\"ummel \& Wagner
(\cite{kue2}) and 
the deep surveys from Metcalfe et al.~(\cite{met95}), Smail et
al.~(\cite{smail}), the Hubble Deep Field (Williams et al.~\cite{wil}),
Hogg et al.~(\cite{hogg}) and the NTT Deep Field (Arnouts et al.~\cite{arn97})
In both filters there is a very good agreement 
between the CADIS data and the average literature counts over the entire
magnitude range. Apart from serving as input for the models presented in
the next section the agreement of our optical data with the
literature confirm that the CADIS-fields contain the typical mix of
galaxies, as it is expected for the wide field of view and the
four lines of sight.

\begin{figure*}
   \resizebox{\hsize}{!}{\includegraphics*[angle=-90]{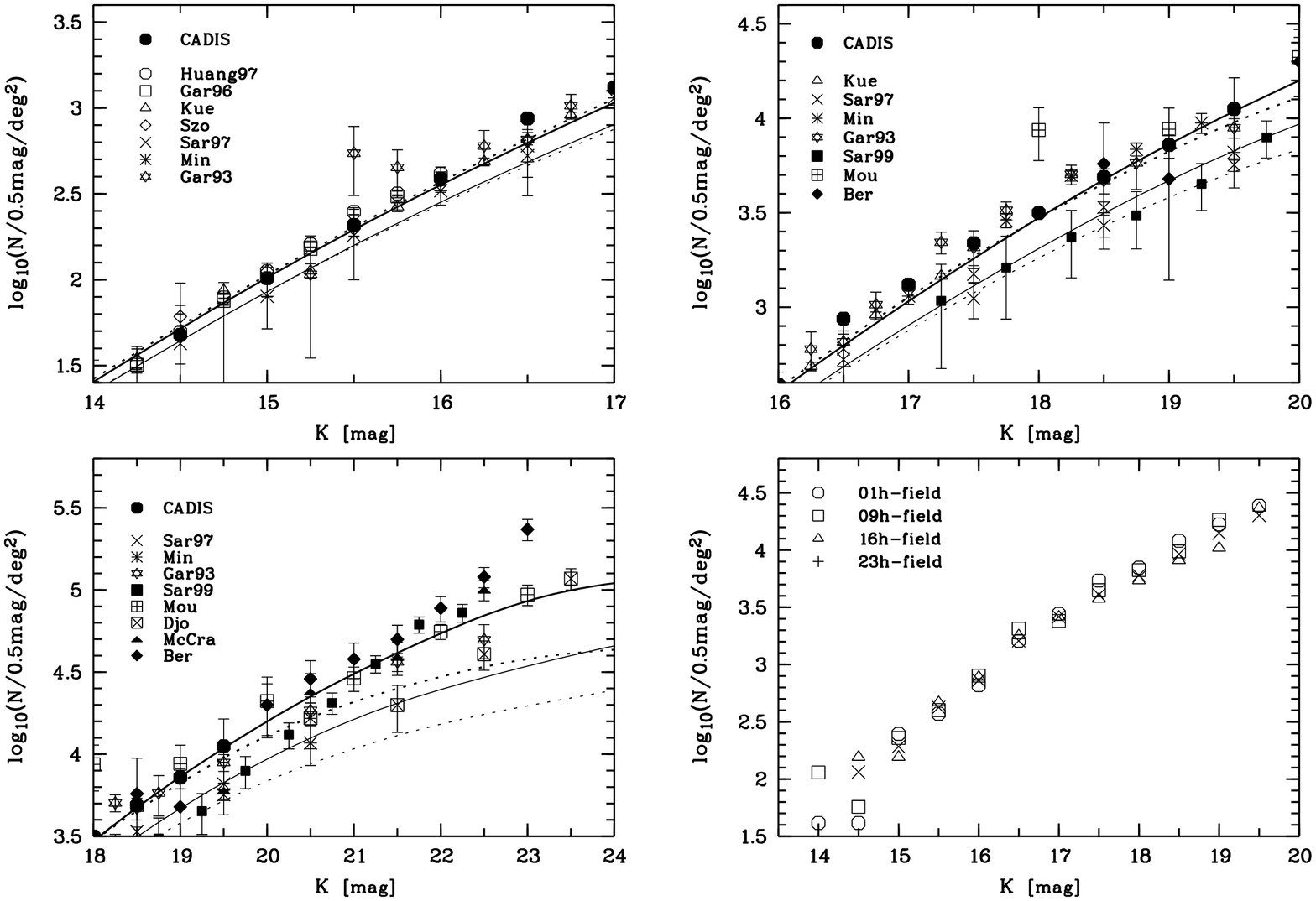}}
   \caption{The $K$-band galaxy number counts.  The filled circles are the 
   number counts obtained in CADIS.  The other symbols are number counts from
   Gardner et al.~(\cite{gar93}) (Gar93),
   Djorgovski et al.~\cite{djo95} (Djo),
   Gardner et al.~(\cite{gar96}) (Gar96),
   Huang et al.~(\cite{hua97}) (Huang97),
   Saracco et al.~(\cite{sar97}) (Sar97),
   Minezaki et al.~\cite{min98}) (Min),
   Moustakas et al.~\cite{mou98} (Mou),
   Bershady et al.~(\cite{bers}) (Ber),
   Szokoly et al., (\cite{szok}) (Szo),
   Saracco et al.~(\cite{sar99}) (Sar99),
   McCracken et al.~(\cite{mcc}) (McCra) and
   K\"ummel \& Wagner (\cite{kue}) (Kue).
   The models for the $K$-band number counts are plotted as in 
   Fig.~\ref{fig:3}.}
   \label{fig:5}
\end{figure*}

The CADIS $K$-band counts are plotted in Fig.~\ref{fig:5}.  We also 
plot for comparison the results from other $K$-band surveys
(
Gardner et al.~\cite{gar93},
Djorgovski et al.~\cite{djo95},
Gardner et al.~\cite{gar96},
Saracco et al.~\cite{sar97},~\cite{sar99},
Huang et al.~\cite{hua97},
Minezaki et al.~\cite{min98},
Moustakas et al.~\cite{mou98},
Bershady et al.~\cite{bers},
Szokoly et al., (\cite{szok}),
McCracken et al.~\cite{mcc} and
K\"ummel \& Wagner \cite{kue}).
Because of the wide range in magnitudes covered by these surveys ($14 < K < 
24$ mag), and the relatively small differences between them, we plot the 
$K$-band number counts in three magnitude ranges for better resolution.\\ 
At magnitudes brighter than
$K=17.5\,\mathrm{mag}$, substantial $K$-band surveys with areas on the
square-degree scale were carried out by
Gardner et al.~(\cite{gar96}), Huang et al.~(\cite{hua97}),
Szokoly et al.~(1999) and K\"ummel \& Wagner (2000). Therefore, the number
counts in this magnitude range are well defined. As shown in the up-left panel
of Fig.~\ref{fig:5}, the CADIS $K$-band number counts in this magnitude range
are in good agreement with previous number counts, but seem to support
only the highest counts in the range $15.5<K<17.5\,\mathrm{mag}$

%
 In the magnitude range of $17.5<K<19.5\,\mathrm{mag}$ (upper right panel of
Fig.~\ref{fig:5}) only the surveys of  Minezaki et
al.~(\cite{min98}, $181\,\mathrm{arcmin}^2$) and Saracco et
al.~(\cite{sar97}, $170\,\mathrm{arcmin}^2$) have a coverage
comparable to CADIS.
With an area of $675\,\mathrm{arcmin}^2$ CADIS is the largest survey in
this magnitude range. The CADIS counts are slightly lower than those of
Minezaki et al.~(\cite{min98}) at the faint end, but $50\,\%$ higher
than the counts of Saracco et al.~(\cite{sar97}) across the
whole brightness interval. Because of the small statistical uncertainties
due to the large $0.2\,\mathrm{deg}^2$ area of the four CADIS fields, our
data provide a more robust measurement of the
number counts in the range $16.5 < K  < 19.5\,\mathrm{mag}$.
The largest remaining discrepancies in the number counts are for
$K>19.5\,\mathrm{mag}$, i.e.~beyond the
limit of the CADIS counts.
At these faint magnitudes the scatter in the counts reach up to a factor of
three (at $K\sim 21.0$, see lower left panel in Fig.~\ref{fig:5}).
From the CADIS counts above the onset of this scatter at
$K\ge 18\,\mathrm{mag}$, we conclude that CADIS favours the largest
values obtained for the number counts at $K\ge 19\,\mathrm{mag}$,
consistent with those obtained by Moustakas et al.~(1997).

  Magnitude errors, field-to-field variation and the Poisson
noise are prime contributors to the uncertainties in the number counts.
Fig.~\ref{fig:5} shows in the lower right panel the number counts for
the individual CADIS-fields. The differential counts for the individual
fields give consistent results and therefore we exclude the possibility
that the average number counts of all fields are affected by field to field
variations. At $K>18\,\mathrm{mag}$ those variations can be significant only
when the sky coverage of a survey is $\ll
 100\,\mathrm{arcmin}^2$.
Accordingly, the field-to-field
variation and the Poisson noise may have significant contribution 
to the scatter of the $K$-band number counts among the deep surveys
which cover only a small area on the order of
$1\,\mathrm{arcmin}^2$.

In most cases, the uncertainties in the number counts result from systematic
errors of the magnitude scale. As it is outlined in Huang et
al.~ (\cite{hua97}), a random magnitude error can cause the slope of the
number counts to become steeper,
while a systematic magnitude error can cause either a shift of number 
counts in the magnitude direction or change of the slope.
The absolute CADIS-photometry (with respect to Vega)
is accurate to $0.1\,\mathrm{mag}$ (Meisenheimer et al. in prep.).
Taking into account those photometric errors leads to perfect agreement
between our counts and the Minezaki et al.~(1998) data, but even then
the CADIS counts are still incompatible with the results of
Saracco et al.~(1997).

%

%

Gardner et al.~(\cite{gar93}) reported finding a break in the logarithmic slope 
of the $K$-band counts at $K \sim 17$, where we have good statistics.  
To test for this, we fit a power law with the form $N(m) = a*10^{b*m}$ 
to our counts in various magnitude ranges.  We find a clear break in 
$\mathrm{d}\log N/\mathrm{d}m$ from a bright-end slope of $b=0.64$ to 
a faint-end slope of $0.36$, with the break in the interval $16.5 < K 
< 17.0\,\mathrm{mag}$.  The slope at the bright end is comparable 
to that found in other surveys (see K\"ummel \& Wagner \cite{kue}).  The 
slope at the faint end is somewhat steeper than values reported in the
smaller surveys of Gardner et al.~(\cite{gar93}) and Minezaki et
al.~(\cite{min98}), who
give slopes around $0.24$.

%
\begin{table}
   \caption{$B$-band number counts \label{tbl:3}}
   \begin{tabular}{cccc}
      \hline
      \noalign{\smallskip}
      B mag& Log(N)\footnotemark& Log(N-$\sigma$)& Log(N+$\sigma$)\\
      \noalign{\smallskip}
      \hline
      \noalign{\smallskip}
      17.0 & 1.07 & 0.54 & 1.30\\    
      18.0 & 1.07 & 0.54 & 1.30\\    
      18.5 & 1.37 & 1.07 & 1.55\\    
      19.0 & 1.62 & 1.41 & 1.76\\    
      19.5 & 2.03 & 1.91 & 2.12\\    
      20.0 & 1.97 & 1.85 & 2.07\\    
      20.5 & 2.44 & 2.37 & 2.50\\    
      21.0 & 2.51 & 2.45 & 2.57\\    
      21.5 & 2.83 & 2.79 & 2.87\\    
      22.0 & 3.07 & 3.04 & 3.10\\    
      22.5 & 3.35 & 3.32 & 3.37\\    
      23.0 & 3.60 & 3.58 & 3.61\\    
      23.5 & 3.85 & 3.83 & 3.86\\    
      24.0 & 4.03 & 4.02 & 4.05\\    
      24.5 & 4.25 & 4.24 & 4.27\\    
      \noalign{\smallskip}
      \hline
   \end{tabular}
\end{table}
\footnotetext{The number counts are in units of 
              $N/0.5\,\mathrm{mag}/\mathrm{deg}^2$.}

%
\begin{table}
   \caption{$R$-band number counts \label{tbl:4}}
   \begin{tabular}{cccc}
      \hline
      \noalign{\smallskip}
      R mag& Log(N)& Log(N-$\sigma$)&Log(N+$\sigma$)\\
      \noalign{\smallskip}
      \hline
      \noalign{\smallskip}
      16.0 &  0.69 &  0.15 &  0.99\\   
      16.5 &  1.39 &  1.13 &  1.55\\   
      17.0 &  1.17 &  0.79 &  1.36\\   
      17.5 &  1.39 &  1.13 &  1.55\\   
      18.0 &  1.87 &  1.74 &  1.96\\   
      18.5 &  2.10 &  2.01 &  2.18\\   
      19.0 &  2.44 &  2.38 &  2.50\\   
      19.5 &  2.68 &  2.63 &  2.72\\   
      20.0 &  2.86 &  2.82 &  2.89\\   
      20.5 &  3.11 &  3.08 &  3.13\\   
      21.0 &  3.27 &  3.25 &  3.30\\   
      21.5 &  3.47 &  3.45 &  3.49\\   
      22.0 &  3.60 &  3.59 &  3.62\\   
      22.5 &  3.78 &  3.77 &  3.80\\   
      23.0 &  4.05 &  4.04 &  4.06\\   
      \noalign{\smallskip}
      \hline
   \end{tabular}
\end{table}

%
\begin{table}
   \caption{$K$-band number counts \label{tbl:5}}
   \begin{tabular}{cccc}
      \hline
      \noalign{\smallskip}
      K mag& Log(N)&Log(N-$\sigma$)&Log(N+$\sigma$)\\
      \noalign{\smallskip}
      \hline
      \noalign{\smallskip}
      14.0 &  1.21 &  0.83 &  1.40\\   
      14.5 &  1.68 &  1.51 &  1.81\\   
      15.0 &  2.01 &  1.90 &  2.10\\   
      15.5 &  2.32 &  2.25 &  2.39\\   
      16.0 &  2.59 &  2.51 &  2.62\\   
      16.5 &  2.94 &  2.91 &  2.97\\   
      17.0 &  3.12 &  3.09 &  3.14\\   
      17.5 &  3.34 &  3.32 &  3.36\\   
      18.0 &  3.50 &  3.48 &  3.51\\   
      18.5 &  3.69 &  3.68 &  3.70\\   
      19.0 &  3.86 &  3.85 &  3.87\\   
      19.5 &  4.05 &  4.04 &  4.06\\   
      \noalign{\smallskip}
      \hline
   \end{tabular}
\end{table}

%
\begin{figure}
   \resizebox{6cm}{!}{\includegraphics*[angle=-0]{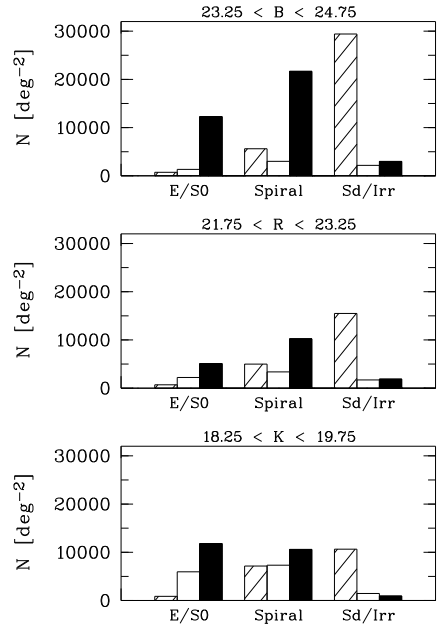}}
   \caption{The observed spectral type mix of galaxies (shaded)
   in the magnitude ranges close to the respective completeness limit
   in comparison to the predicted type-mix of the no-evolution and the
   passive-evolution model (the white and black bins, respectively).}
   \label{fig:8}
\end{figure}

\section{Modeling the Number Counts and Galaxy Population} \label{sec:mod}

Models of galaxy number counts are primarily constrained by assumptions 
on the cosmological geometry, galaxy spectral evolution, and the galaxy 
luminosity function (see Gardner \cite{gardner98}).  Theoretical uncertainties 
on the number count models come mainly from modeling the spectral evolution 
of galaxies, especially in the infrared bands.  Many galaxy spectral synthesis 
models currently exist, ranging from simple stellar population models assuming 
an instantaneous burst of star formation to models including gas, recycling of 
metals, and nebular emission (Bruzual \& Charlot~\cite{bru93},
Bertelli et al.~\cite{bet94}, Worthey~\cite{wor94},
Fioc \& Rocca-Volmerange~\cite{fio97}).

Most models make predictions consistent with each other at visible wavelengths 
(Charlot \cite{cha96}).  There is, however, a significant scatter between 
model predictions at near-infrared wavelengths.  This is due primarily to 
poor knowledge of the atmospheric parameters for cool stars and a substantial 
contribution from the nebular emission (Fioc \& Rocca-Volmerange \cite{fio97}).  
Many number count models (Metcalfe et al.~\cite{met96}, McCracken et 
al.~\cite{mcc}) use simple stellar population models convolved with various 
star formation histories (Bruzual \& Charlot \cite{bru93}), which over-predict 
the number counts at bright magnitudes at both visible and near-infrared 
wavelengths.  We use PEGASE 
({\em Projet d'Etude des Galaxies par Synthese Evolutive}), a new 
spectrophotometric evolution model which includes gas and dust components, 
nebular emission, and a better description of galaxy SEDs in the 
near-infrared (Fioc \& Rocca-Volmerange~\cite{fio97}).  Our number
count model constructed using PEGASE is able to fit the observed number counts
at bright magnitudes well (Fioc \& Rocca-Volmerange \cite{fio97}).  

We adopt the passive evolution and no-evolution number counts models from 
Huang (\cite{hua97}).  The cosmological k-correction and evolution 
e-correction are calculated using the PEGASE model.  We plot model predictions 
against the observed number counts in Figs.~\ref{fig:2}-\ref{fig:5}. 
Huang (\cite{hua97}) has shown that these models correctly predict the 
surface density and color distribution of local galaxy populations, consistent 
with Fioc \& Rocca-Volmerange (\cite{fio97}).  The model Hubble types refer to the 
spectral type and not the morphological type, although they are correlated.  
The spectral types are defined by the e-folding time of the star formation 
rate.  Most models use an exponentially decreasing star formation rate for 
all spectral types except Irr galaxies, which are so blue that the star 
formation rate must have been constant or increasing over the $\sim$4.5\,Gyr 
life of the galaxy. 

%
\begin{figure}
   \resizebox{\hsize}{!}{\includegraphics*[angle=-0]{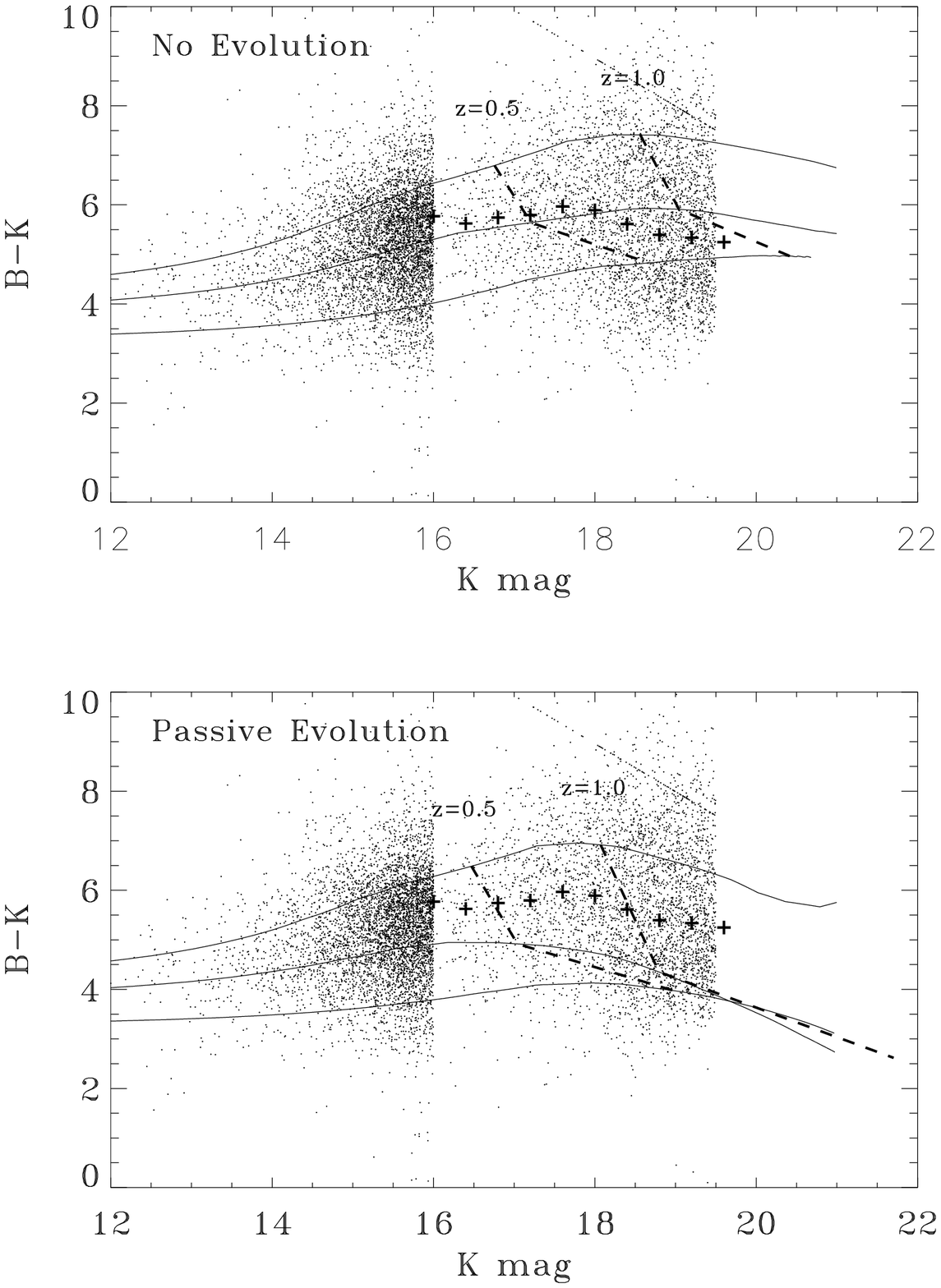}}
   \caption{$B-K$ vs.~ $K$ for the $K$-selected samples.  The bright 
   $K$-selected  sample ($K<16\,\mathrm{mag}$) is obtained by Huang et 
   al.~(\cite{hua97}). The no-evolution and passive evolution models are plotted 
   in separate panels. In each panel, the upper solid line is the 
   $1.5\,L_*$ elliptical galaxy model; the middle line is the $L_*$ Sbc 
   galaxy model; and the lower line is the $0.25\,L_*$ Sd/Irr galaxy model. 
   The dashed lines link the colors among the different models at $z=0.5$ 
   and $z=1.0$ (for $h_0 =0.5$, $q_0=0.05$).  Though the dashed lines are 
   not exactly redshift contours, they roughly outline a redshift distribution 
   on the color-magnitude relation. The cross is a medium $B-K$ color at each 
   magnitude bin.}
   \label{fig:6}
\end{figure}

%
\begin{figure}
   \resizebox{\hsize}{!}{\includegraphics*[angle=-0]{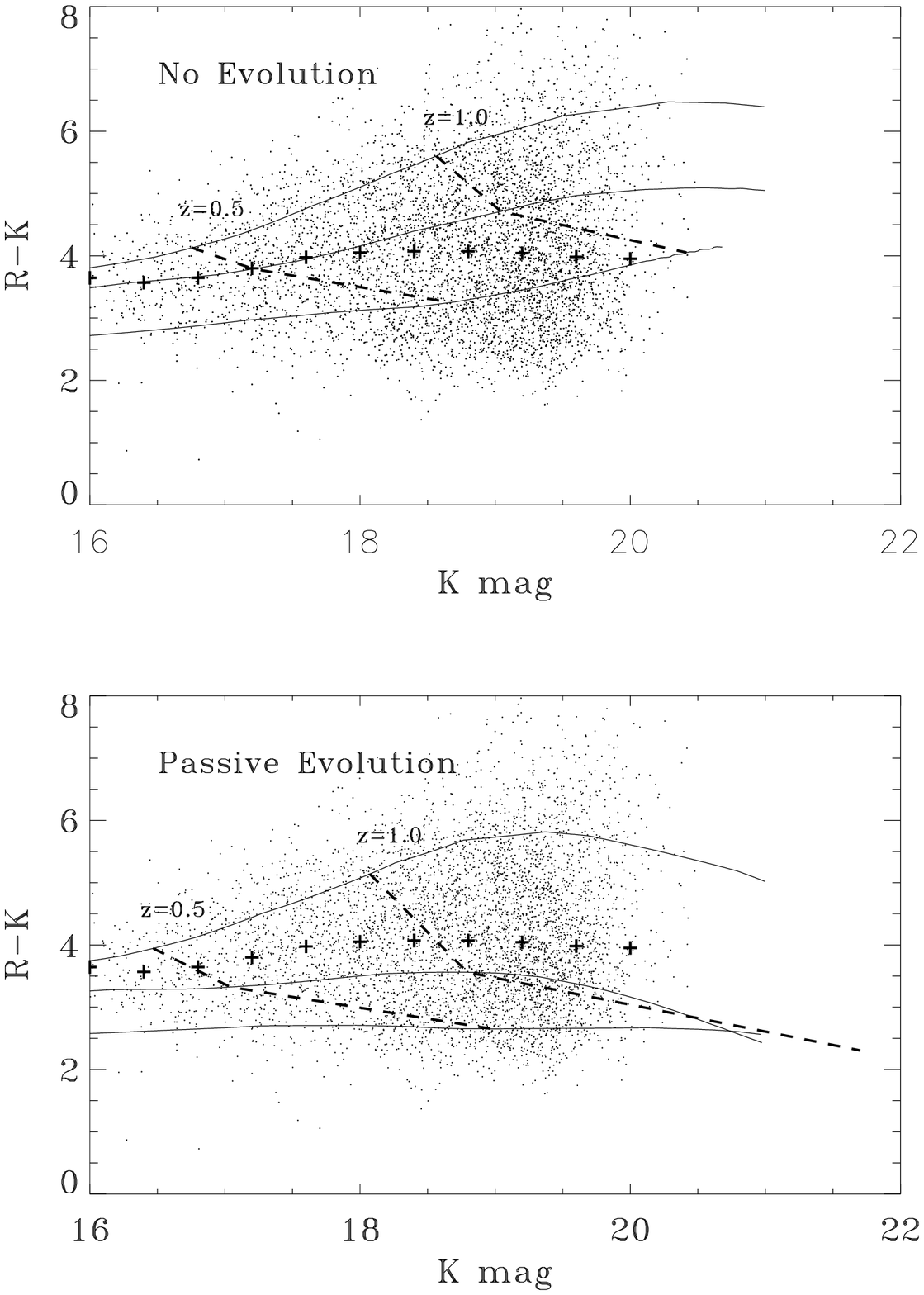}}
   \caption{$R-K$ vs.~ $K$ for the $K$-selected samples.  The models and the 
   medium colors are plotted in the same way as in Fig.\ \ref{fig:6}.}
   \label{fig:7}
\end{figure}

We can see in Figs.~\ref{fig:3}-\ref{fig:5} that evolution has the strongest 
effect on the models, while the geometry ($q_0=0.05$ and $0.5$) does not make 
a large contribution at brighter magnitudes ($B < 24\,\mathrm{mag}$, $R < 
23\,\mathrm{mag}$ and $K < 20\,\mathrm{mag}$).  We adopt a high normalization 
of $\phi_* = 0.02\,\mathrm{h}^3 \mathrm{Mpc}^{-3}$.  Only the passive evolution
model is able to simultaneously fit the $B$-, $R$- and  $K$-band number counts 
in a low density universe reasonably well.  The no-evolution models are 
significantly flatter than the observational data.  The $R$-band number counts
show the highest excess and may favor an even higher normalization.  The 
$K$-band counts have the lowest excess, but the no-evolution model still 
significantly under-predicts the counts for $K > 16\,\mathrm{mag}$.

We note that Metcalfe et al.\ (\cite{met96}) and McCracken et al.\ (\cite{mcc})
are able 
to fit their no-evolution model to the observed $K$-band number counts at 
faint magnitudes, while their passive evolution model predicts more galaxies 
than observed.  This discrepancy between our model and theirs is due to the 
differences in the galaxy spectral evolution models in the $K$-band.
  
Although our passively evolving model is able to fit the observed number 
counts in the optical and near-infrared bands well, this does not necessarily
mean that the passive evolution model is a correct physical description
for galaxy evolution at intermediate redshifts.  This is rather evident in 
Fig.~\ref{fig:8} where we compare the spectral mix of our galaxies,
as determined from their SEDs, with the predicted type mix from the
no-evolution and the passive-evolution models.  In all three bands this
comparison is done in the $1.5\,\mathrm{mag}$ interval just brighter than
the completeness limit of our data, where the statistics are best.
The data and model predictions obviously disagree.  The passive evolution 
model predicts that early type galaxies are mainly responsible for
the excess over no-evolution model at faint magnitudes and high redshifts, 
Stellar evolution would make these early type galaxies much brighter
at high redshifts because their stellar populations were younger.  However, 
the spectral type mix of our number counts shows that even in the $K$-band 
spirals and SD/Irr are by far the main contributors to the counts.  This can 
be understood if a population of star forming dwarf galaxies at $z < 1.0$ 
dominates the number counts at faint magnitudes.

Another way of studying galaxy populations and their evolution is
through color-magnitude diagrams.  Fig.~\ref{fig:6} ($B-K$ vs.~ $K$) 
and Fig.~\ref{fig:7} ($R-K$ vs.~ $K$) are the color magnitude diagrams 
for the $K$-selected sample.  The $B-K$ colors for the $K$-selected bright 
sample obtained by Huang (\cite{hua97}) are also plotted in Fig.~\ref{fig:6}.
To compare with the data, we derived the $B-K$ and $R-K$ colors as
a function of apparent magnitude $K$ for a $1.5\,L_*$ elliptical galaxy,
a $L_*$ Sbc galaxy and a $0.25\,L_*$ Sd/Irr galaxy and plot them together
with the observed color-magnitude data. Those three galaxy types have been
used to calculate the model counts presented above, and they represent
the populations with the reddest, medium, and bluest colors
(the $B-K$ and $R-K$ colors at $z=0$ are given in Table~\ref{tbl:6}).
Provided the models accurately represent the color evolution of galaxies, 
the E and Sd/Irr model colors should bracket the observed distribution of 
colors and Sbc model colors trace the median of the color distribution.  As 
seen in Fig.~\ref{fig:6}, the no-evolution and passive evolution models 
both bracket the $B-K$ color distribution at bright magnitudes very well, 
and both predict a break in the $B-K$ color at $K \sim 18\,\mathrm{mag}$.  
For the bright sample of Huang (\cite{hua97}), there is no significant 
difference between the no-evolution and passive evolution models.  However, 
at $K > 18\,\mathrm{mag}$ the non evolving Sd/Irr galaxy model 
colors are redder than a significant fraction of the observed colors.
The passive evolution model predicts much bluer colors for these
three types of galaxies at $K > 18\,\mathrm{mag}$ and gives a better
representation of the observed color distribution.

The dashed lines in Figs.~\ref{fig:6} and \ref{fig:7} link the different 
models at $z=0.5$ and $z=1.0$ (for $h_0 = 0.5$, $q_0=0.05$).  We can see 
why even the passive evolution model was unable to predict the type mix
in Fig.~\ref{fig:8}: the ellipticals were brighter in the past and thus 
visible out to higher redshifts ($z>1$), making a significant contribution 
to the faint number counts.  There are two obvious ways to explain this 
discrepancy between the observations and the models:\\
(i) Massive galaxies with high star formation rates at $z>1$ appear bluer 
and are accordingly classified as SED type later than E/S0, even though 
they will end up as early type galaxies at $z=0$.\\
(ii) less massive galaxies do not undergo significant star formation 
until relatively late times ($z < 1$).\\
The good statistics from large galaxy samples and the spectral type studies 
favor the second alternative.  However, it will be necessary to compare the 
model spectral type with the actual morphological Hubble type of the $K > 
18.0\,\mathrm{mag}$ galaxies before this question can be settled.

%
\begin{table}
   \caption{Colors for E/S0, Sbc and Sd/Irr at z=0\label{tbl:6}}
   \begin{tabular}{cccc}
      \noalign{\smallskip}
      \hline
      \noalign{\smallskip}
             & E/S0 & Sbc  & Sd/Irr \\
      $B-K$  & 4.18 & 3.86 & 3.28   \\
      $R-K$  & 2.62 & 2.63 & 2.30   \\
      \noalign{\smallskip}
      \hline
   \end{tabular}
\end{table}

\section{Summary} \label{sec:con}

  We derived the $B$-, $R$- and $K$-band number counts in the Calar Alto
Deep Imaging Survey. The total coverage for the $K$-band imaging is about
$0.2\,\mathrm{deg}^2$. The $B$-band number counts cover the range from
$B=16.75\,\mathrm{mag}$ to $B=24.75\,\mathrm{mag}$; the $R$-band
number counts cover from $R=15.75\,\mathrm{mag}$ to $R=23.25\,\mathrm{mag}$; 
and the $K$-band number counts cover
from $K=13.75\,\mathrm{mag}$ to $K=19.75\,\mathrm{mag}$.\\
Our number counts at optical wavelength match the number counts of
other surveys very well. Our $K$-band counts differ from the results of
previous surveys by up to $50\,\%$. Because of the large area
covered by our survey down to $K=19.75\,\mathrm{mag}$, the CADIS number counts
in $K$ set a new standard in the range $17.75<K<19.75\,\mathrm{mag}$,
where our survey represents the largest $K$-band survey to date.
Between $K< 16.5\,\mathrm{mag}$ and $K> 17.5\,\mathrm{mag}$ our counts
display a break in the logarithmic slope from $0.64$ to $0.36$, 
we estimate this break occurs around $K=16.7\,\mathrm{mag}$.

In order to get a first insight in the underlying evolution process
we also adopt no-evolution and passive evolution models to
fit our optical and near-infrared number counts. We
find that both our optical and near-infrared number counts have an excess
over the no-evolution model and can be simultaneously well fitted
by the passive evolution model. The passive evolution 
model, however, predicts that the excess of the
number counts at faint magnitudes is mainly due to the early type galaxies
which were brighter at high redshifts. This is in contradiction to the
observed spectral type mix contributions to our counts, which seems to be
dominated by spirals and star-forming galaxies
down to the faintest limits of our survey.\\
Thus we conclude that both non-evolving and passive evolution are both
in contradiction with our observed number counts and the galaxy type mix at the
faint magnitude end of our survey.
The most sensible explanation for this finding would be a population
at $z\le 1.0$ with violent star burst events. These make them very prominent
at intermediate redshifts while locally (at $z\sim 0$) they appear at such
faint ends of both, the total luminosity and surface brightness range
that they are underrepresented in current redshift surveys.


\begin{thebibliography}{}
\bibitem[1996] {abr96}{Abraham R.G., Tanvir N.R., Santiago B.X., et al.,
        1996, MNRAS, 279, L47}
\bibitem[1997] {arn97}{Arnouts S., de Lapparent V., Mathez G., et al.,
        1997, A\&A Supl., 124, 163}
\bibitem[1999] {arn99}{Arnouts S., D'Odorico S., Cristiani S., et al.,
        1999, A\&A, 341, 641}
\bibitem[1998] {bers}{Bershady M.A., Lowenthal J.D., Koo D.C., 1998, ApJ,
        505, 50}
\bibitem[1996] {ber96}{Bertin E., Arnouts S., 1996, A\&A, 117, 393
\bibitem[1994] {bet94}{Bertelli G., Bressan A., Chiosi C., Fagotto F., 
        Nasi E., 1994, A\&AS, 106, 275}
\bibitem[1998] {biz98}{Bizenberger P., McCaughrean M., Thompson D., Birk C.,
        1998, SPIE Conference Proceedings. vol. 3354, 109}
\bibitem[1993] {bru93}{Bruzual G., Charlot S. 1993, ApJ, 405, 538}
\bibitem[1992] {cas92}{Casali M.M., Hawarden T.,  1992, {\it JCMT UKIRT 
        Newsletter}, 4, 33}
\bibitem[1996] {cha96}{Charlot S., 1996, in From Stars to Galaxies: 
        The Impact of Stellar Physics on Galaxy Evolution, ed.~ C. Leitherer, 
        U. Fritze-von-Alvensleben, and J. Huchra, ASP Conference Series
        Vol.\ 98}
\bibitem[1994] {cow94}{Cowie L.L., Gardner J.P., Hu E.M., et al.,
        1994, ApJ, 434, 114}
\bibitem[1996]{cow96}{Cowie L.L.,  Songaila A., Hu E.M. Cohen J.G., 1996,
        AJ, 112, 839}
\bibitem[1997]{cow97}{Cowie L.L.,  Hu E.M., Songaila A., Egami, E., 1997,
        ApJ, 481, L9}
\bibitem[2000]{dic00}}{Dickinson M., 2000, in Building Galaxies: 
        From the Primordial Universe to the Present, 
        ed.~ F. Hammer, T.X. Thuan, V. Cayatte, B. Guiderdoni, 
        \& J. Tranh Than Van, (Paris:Ed. Frontieres)}
\bibitem[1995] {djo95}{Djorgovski S., Soifer B.T., Pahre M.A., et al.,
        1995, ApJ, 438, L13}
\bibitem[1996]{ell96}{Ellis R.S., Colless M., Broadhurst T.J., 
        Heyl J.S., Glazebrook K., 1996, MNRAS, 280, 235}
\bibitem[1997] {ell97}{Ellis R.S., 1997, ARAA, 35, 389}
\bibitem[1997] {fio97}{Fioc M., Rocca-Volmerange B., 1997, A\&A, 326, 950}
\bibitem[2001] {fried}{Fried J.W., von Kuhlmann B., Meisenheimer W., et al.,
        2001, A\&A, accepted (astrph/0012343)}
\bibitem[1996] {gar96}{Gardner J.P., Sharples R.M., Carrasco B.E.,
        Frenk C.S., 1996, MNRAS, 282, L1}
\bibitem[1998] {gardner98}{Gardner J.P., 1998. PASP 110,291}
\bibitem[1993] {gar93}{Gardner J.P., Cowie L.L., Wainscoat R.J.,
        1993, ApJ, 415 L9}
\bibitem[1995]{gla95}{Glazebrook K., Ellis R.S., Santiago B.X., 
        Griffiths R., 1995, MNRAS, 275, L19}
\bibitem[1997]{hogg}{Hogg D.W., Pahre M.A., McCarthy J.K., et al.,
        1997, MNRAS 288, 404}
\bibitem[1997] {hua97}{Huang J.-S., Cowie L.L., Gardner J.P., et al.,
        1997, ApJ, 476, 12}
\bibitem[1997] {hua97t}{Huang J.-S., 1997, Ph.D. Thesis, University of Hawaii}
\bibitem[1998] {hua98}{Huang J.-S., Cowie L.L., Luppino G.A., 1998, ApJ, 
        496, 31}
\bibitem[1996] {kin}{Kinney A.L., Calzetti D.A., Bohlin R.C., et al.,
        1996, ApJ, 467, 38}
\bibitem[2000] {kue}{K\"ummel M.W., Wagner S.J., 2000, A\&A, 353, 867} 
\bibitem[2001] {kue2}{K\"ummel M.W., Wagner S.J., 2001, A\&A, submitted} 
\bibitem[1995] {lil95}{Lilly S.J., Tresse L., Hammer F., Crampton D.,
        LeF\`evre O., 1995, ApJ, 455, 108}
\bibitem[2000] {mcc} {McCracken H.J., Metcalfe N., Shanks T., et al.,
        2000, MNRAS, 311, 707}
\bibitem[1986] {mei86}{Meisenheimer K.,  R\"oser H.-J., 1986, in 
        The Optimization of the Use of CCD Detektors in Astronomy, 
        ed.~ J.-P.~Baluteau \& S.~D'Odorico (Munich:ESO-OHP), 227}
\bibitem[1998] {mei99}{Meisenheimer K., Beckwith S., Fockenbrock R., 
        et al., 1998, in ASP Conf. Ser. 146, in The Young Universe, 
        ed S.~D'Odorico, A.~Fontana, \& E.~Giallongo (San Francisco:ASP), 134}
\bibitem[1995] {met95}{Metcalfe N., Shanks I., Fong R., Roche N.,
        1995, MNRAS, 273, 257}
\bibitem[1996] {met96}{Metcalfe N., Shanks I., Campos R., Fong R., Gardner
        J. P., 1996, Nature, 383, 236}
\bibitem[1998] {min98}{Minezaki T., Kobayashii Y., Yoshii Y.,
        Peterson B.A., 1998, ApJ, 494, 111}
\bibitem[1998] {mou98}{Moustakas L.A., Davis M., Graham J.R., et al.,
        1997, ApJ, 475, 445}
\bibitem[1999] {pra} {Prandoni I., Wichmann R., da Costa L., et al.,
        1999, A\&A, 345, 448}
\bibitem[1997] {sar97}{Saracco P., Iovino A., Garilli B., Maccagni D.,
        Chincarini, C., 1997, AJ, 114, 887}
\bibitem[1999] {sar99}{Saracco P., D'Odorico S., Moorwood A., et al,
        1999, A\&A, 349, 751}
\bibitem[1995] {smail}{Smail I., Hogg D.W., Yan L., Cohen J.G., 1995,
        ApJ 449, L105}
\bibitem[1994] {son94}{Songaila A., Cowie L.L., Hu E.M., Gardner J.P.,
        1994, ApJS, 94, 461}
\bibitem[1996] {ste96}{Steidel, C. C., Giavalisco, M., Petini, M.,
        Dickinson, M., Adelberger, K. L., 1996 ApJ, 462, L17}
\bibitem[1999] {ste99}{Steidel, C. C., Adelberger, K. L., Giavalisco, M.,
        Dickinson, M., Petini, M., 1999, ApJ, 519, 1}
\bibitem[1999]{szok}{Szokoly G.P., Subbarao M.U., Connolly A.J.,
        Mobasher B., 1999, ApJ, 492, 452}
\bibitem[1998] {tho98}{Thommes E., Meisenheimer,K., Fockenbrock R., et al.,
        1998, MNRAS, 293, L6}
\bibitem[1999] {tho99}{Thompson D., Beckwith S.V.W., Fockenbrock R., et al.,
        1999, ApJ, 523, 100}
\bibitem[1996] {vdb96}{van den Bergh S., Abraham R.G., Ellis R.S., et al.,
        1996, AJ, 112, 359}
\bibitem[1992] {wai92}{Wainscoat R.J., Cowie L.L., 1992, AJ, 103, 332}
\bibitem[1996] {wil} {Williams R.E., Blacker B., Dickinson M., et al., 
        1996, AJ 112, 1335}
\bibitem[1994] {wor94}{Worthy G., 1994, ApJS, 101, 181}
\bibitem[1998] {wo98}{Wolf C., 1998, Ph.D. Thesis, University of Heidelberg}
\bibitem[1999] {wo99}{Wolf C., Mundt R., Thompson D., et al.,
        1999, A\&A, 343, 399}
\end{thebibliography}
\end{document}